\documentclass{pasj01}
\usepackage{lineno}
\draft 
\Received{$\langle$reception date$\rangle$}
\Accepted{$\langle$acception date$\rangle$}
\Published{$\langle$publication date$\rangle$}
\usepackage{wrapfig}
\usepackage{color}
\usepackage{soul}
\usepackage{ulem}

\begin{document}

\title{Magnetic activity variability from H$\alpha$ line intensive monitoring for two F-type stars having a hot-Jupiter, $\tau$ Bootis A and $\upsilon$ Andromedae A}
\author{
 Sanghee \textsc{Lee},\altaffilmark{1}
 Yuta \textsc{Notsu},\altaffilmark{1,2,3}
 and
 Bun'ei \textsc{Sato}\altaffilmark{1}
 }
 
   \altaffiltext{1}{Department of Earth and Planetary Sciences, Tokyo Institute of Technology,
   2-12-1 Ookayama, Meguro-ku, Tokyo 152-8551, Japan}
 \altaffiltext{2}{Laboratory for Atmospheric and Space Physics, University of Colorado Boulder, 3665 Discovery Drive, Boulder, CO 80303, USA}
 \altaffiltext{3}{National Solar Observatory, 3665 Discovery Drive, Boulder, CO 80303, USA}

\email{lee.s.ay@m.titech.ac.jp}


\maketitle

\begin{abstract}
We report the results of intensive monitoring of the variability in the H$\alpha$ line for two F-type stars, $\tau$ Boo and $\upsilon$ And, during the last four years 2019-2022, in order to investigate their stellar magnetic activity. The 4-year H$\alpha$ line intensity data taken with the 1.88-m reflector at Okayama Branch Office, Subaru Telescope, shows the existence of a possible $\sim$ 123-day magnetic activity cycle of $\tau$ Boo. The result of the H$\alpha$ variability as another tracer of the magnetic activity on the chromosphere is consistent with previous studies of the Ca II H\&K line and suggests that the magnetic activity cycle is persisted in $\tau$ Boo. For $\upsilon$ And, we suggest a quadratic long-term trend in the H$\alpha$ variability.
Meanwhile, the short-term monitoring shows no significant period corresponding to specific variations likely induced by their hot-Jupiter in both cases ($\approx$ 3.31 and 4.62 days, respectively). In this H$\alpha$ observation, we could not find any signature of the Star-Planet Magnetic Interaction. It is speculated that the detected magnetic activity variability of the two F-type stars is related to the stellar intrinsic dynamo.     
\end{abstract}

\section{Introduction}
In the Sun-like stars, the atmosphere exhibits various phenomena, generally induced by the magnetic fields and the active regions in the stellar convective outer atmosphere. The stellar magnetic activity is observed on the outer atmosphere, the chromosphere, where we could observe emission lines in some specific species. The chromospheric emission is observed in various spectral lines such as Ca II H\&K, H$\alpha$ and H$\beta$ line, especially the Ca II H\&K line cores (3933, 3968 \AA) are most well-studied to diagnose the stellar magnetic activity (\cite{linsky:2019}).

The chromospheric lines monitoring on a long time-scale has revealed stellar magnetic activity cycles for the Sun-like stars. The magnetic activity cycle is important to understand the stellar magnetic activity behaviors. The magnetic activity of stars having a convective zone is related to the stellar dynamo (\cite{piddington:1982}) and therefore the study of the magnetic activity cycle leads us to understand the stellar magnetic field and the dynamo. 
For the Sun, the Ca II H\&K line core emission is caused by plages on its chromosphere, and it varies over the 22-yr solar magnetic activity cycle (11-yr sunspot cycle). \citet{baliunas:1995} reported the magnetic activity cycle for 46 among 112 main-sequence stars of spectral type F2-M2 from the Mount Wilson HK program over several decades. Most G-type and K-type stars of the samples showed typical cycle periods varying from 2.5 yr to more than 20 yr (see also \cite{metcalfe:2017}). As a result of placing the Mount Wilson samples in the HR diagram, the activity of G- and K-type stars have long and pronounced cycles (\cite{schroder:2013}). However, the solar and the stellar dynamo are still poorly understood despite the continuous study of the solar cycle discovered many decades ago (\cite{charbonneau:2020}).   

Meanwhile, F-type stars also have a convective zone so that their magnetic activity is driven by the dynamo. For F-type stars, although there have not been so many cases to suggest their activity cycle (\cite{baliunas:1995}; \cite{brandenburg:2017}), recent research suggested that F-type stars have a relatively short magnetic activity cycle. \citet{metcalfe:2010} and \citet{mittag:2019} supposed that some F-type stars show a periodic variation within 1-2 years. This short-term magnetic activity cycle could provide an ideal time-frame to study the stellar dynamo. Thus, it is important to increase observations of F-type stars.

The monitoring of chromospheric lines has also been performed, focusing on the planetary perspective. An extremely close-in giant planet, a hot-Jupiter, can influence the magnetic activity on the chromosphere of its parent star through a magnetic reconnection between the stellar and the planetary magnetic field (Star-Planet Magnetic Interaction; SPMI) (\cite{cuntz:2000}; \cite{rubenstein:2000}) or a tidal effect (\cite{cuntz:2000}; section 6 of \cite{shibata:2013}).
A few systems having a hot Jupiter are known to show an activity enhancement along the planetary orbital phase by observing the chromospheric lines (\cite{shkolnik:2008}). The SPMI signature has been studied to explore various influences on the stars and hot-Jupiters (\cite{cauley:2019}). The dynamo could also be accelerated by interactions between the star and its hot-Jupiter (\cite{cuntz:2000}).
More robust spectroscopic evidence of the SPMI will be needed to clarify its effects on the stellar magnetic activity. 

In this paper, we present long-term monitoring of two F-type stars having a hot-Jupiter, $\tau$ Boo and $\upsilon$ And, using the H$\alpha$ line observed with the 1.88-m reflector at Okayama Branch office, Subaru Telescope, to investigate their magnetic activity behaviors. The stellar parameters for these two target stars are summarized in Table \ref{tbl1}. 

$\tau$ Bootis A (HD120136, F7V) was identified as a hot-Jupiter system (M$_p$ sin \textit{i} = 4.4 M$_J$, P$_{orb}$ = 3.31 d, a = 0.0462 AU) (\cite{butler:1997}). \citet{mittag:2017} found a 122-day variation in the Ca II H\&K data for the star, consistent with other results from their individual Ca II H\&K data (\cite{baliunas:1997}; \cite{mengel:2016}). In addition, using spectropolarimetry, \citet{fares:2013} suggested that the polarity reversal of its magnetic field shows a possible magnetic cycle of 240 or 740 days, which would correspond to a chromospheric cycle (i.e. the magnetic activity cycle on the chromosphere) period of 120 or 370 days. The possible magnetic cycle of 240 days is precisely accordance with the chromospheric cycle of 122 days from \citet{mittag:2017}. Also, the SPMI in $\tau$ Boo was suggested by a Ca II H\&K variation synchronized with the planetary orbital period (\cite{shkolnik:2008}). However, its activity behaviors remain some questions. The results from \citet{fares:2013} do not rule out the possibility of 370 days for the stellar magnetic activity cycle. Additionally, \citet{mittag:2017} assumed that there is a shift of phase in the magnetic activity variability, which might imply any other activity behaviors of $\tau$ Boo.

$\upsilon$ Andromedae A (HD9826, F8V) has a multi-planet system and the innermost planet is a hot-Jupiter (M$_p$ sin \textit{i} = 1.70 M$_J$, P$_{orb}$ = 4.62 d, a = 0.0594 AU) (\cite{butler:1997}). The previous Ca II H\&K results of $\upsilon$ And suggested the existence of the SPMI signature (\cite{shkolnik:2005}). However, to date, there have been no reports of the magnetic activity cycles detected by long-term monitoring of the chromospheric lines. 

The rest of this paper is organized as follows. The observations and data reduction are described in Section 2. Analyses and results of the observed H$\alpha$ line are presented in Section 3, and we discuss the results in Section 4. Section 5 is devoted to a summary.

\begin{table}[t]
\caption{Basic parameters of the target stars $\tau$ Boo A and $\upsilon$ And A, and hot-Jupiters $\tau$ Boo Ab and $\upsilon$ And Ab}\label{tbl1}
\begin{center}
\begin{tabular}{ccccccccccc}\hline\hline
Parameter      & $\tau$ Boo A & $\upsilon$ And A & Reference\\
\hline			   			   
Spectral Type         & F7V        & F8V &\\
$V$ (mag)             & 4.50   &  4.10 & 1 \\
$T_{\rm eff}$ (K)  & 6399$\pm$45  & 6213$\pm$44 & 2,3\\ 
$\log g$ (cgs)   & 4.27$\pm$0.06 & 4.0$\pm$0.1 & 2,3\\
$Age$ (Gyr) & 0.9$\pm$0.5 & 3.12$\pm$0.12 & 2,4\\
$R$ ($R_{\odot}$) & 1.42$\pm$0.08 & 1.480$\pm$0.087  & 2,5\\
$M$ ($M_{\odot}$) & 1.39$\pm$0.25 & 1.27$\pm$0.06 & 2,4 \\
$v\sin i$ (km s$^{-1}$) & 14.27$\pm$0.06 & 9.5$\pm$0.8 & 2,4 \\
$P_{rot}$ (day) & 3.1$\pm$0.1 & 7.3$\pm$0.04 & 6,7\\
$P_{orb}$ (day) & 3.31 & 4.62$\pm$0.23 & 8,9\\
$M_{p}$ sin \textit{i} (M$_J$) &  4.4 & 1.70 & 10  \\  
$a$  (AU) &  0.0462 & 0.0594 & 10  \\  
                & ($\tau$ Boo b) & ($\upsilon$ And b)   &    \\

\hline
\end{tabular}
\end{center}
$^1$SIMBAD, $^2$\,\citet{borsa:2015}, $^3$\,\citet{fuhrmann:1998}, $^4$\,\citet{mcarthur:2010}, $^5$\,\citet{vanbelle:2009}, $^6$\,\citet{mengel:2016}, $^7$\,\citet{simpson:2010}, $^8$\,\citet{butler:2006}, $^9$\,\citet{piskorz:2017}, $^{10}$\,\citet{butler:1997}

\end{table}

\section{Observation and data reduction}

\subsection{HIDES-F Observation}
We observed spectroscopic data of these two target stars using the HIgh Dispersion Echelle Spectrograph (HIDES; \cite{izumiura:1999}) of the 1.88-m reflector at Okayama Branch Office, Subaru Telescope. The three CCDs of HIDES cover a wavelength region of 3700-7500 \AA. The spectra were obtained by Fiber-Feed mode (HIDES-F; \cite{kambe:2013}), and a high-efficiency mode with an image slicer was adopted. The HIDES was upgraded in 2018. In this upgrade, the slit mode elements were removed from HIDES optical path, and optical instruments were rearranged on a new stabilized platform in the precise temperature-controlled Coud\'e room to enhance the performance.
For our analysis, $\tau$ Boo was observed on 173 nights, and $\upsilon$ And was observed on 84 nights with HIDES-F between March 2019 and June 2022. The width of the sliced image was $\timeform{1.05''}$ corresponding to the spectral resolution of \textit{R} $\sim$ 55000 by about 3.8-pixel sampling. We obtained the signal-to-noise ratio of S$/$N $\sim$ 100-150 per pixel at 6563 \AA\ within an exposure time of 700 seconds for the stars (slightly different depending on weather conditions). For HIDES-F, there were severe aperture overlaps among 3700-4000 \AA\ owing to the use of the image slicer, therefore the scattered light for these apertures could not be subtracted. Consequently, we discard these overlapped apertures and did not analyze the Ca II H lines for HIDES-F spectra.

\subsection{Echelle data reduction and telluric correction}
The echelle data reduction was performed in a standard manner using the IRAF\footnote{IRAF is distributed by the National Optical Astronomy Observatories, which is operated by the Association of Universities for Research in Astronomy, Inc. under a cooperative agreement with the National Science Foundation, USA} packages to extract one-dimensional raw spectra (bias subtraction, flat fielding, scattered-light subtraction and spectrum extraction). The wave-length calibration was determined from Thorium-Argon arcs taken before and after each spectrum. 
For each night spectrum, we carried out the correction for telluric absorption lines using $\alpha$ Leo (B8IV, \textit{v}sin \textit{i} = 317 km s$^{-1}$), which was taken on the same night as a standard star. This star is rapidly rotating so that it has broadened, featureless stellar spectra and provides only telluric lines. We can use the spectrum of $\alpha$ Leo for the telluric line correction. We also corrected the dependence on airmass, which varies with the elevation of the object, by using the standard star spectrum (\cite{kawauchi:2018}).

\begin{figure}[h]
 \begin{center}
  \includegraphics[width=68mm]{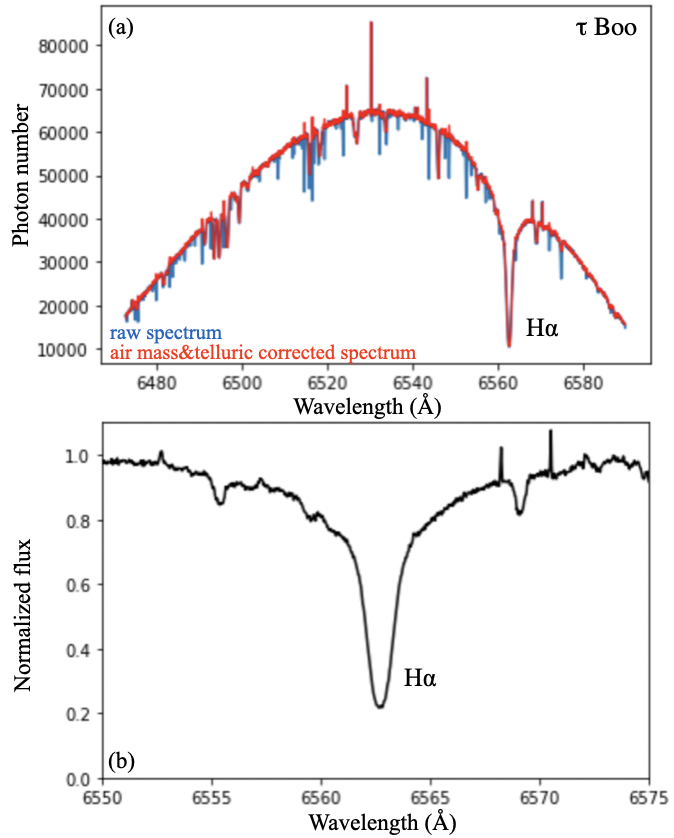}
 \end{center}
 \caption{(a) Single raw spectrum (blue) and telluric line corrected spectrum (red) of H$\alpha$ line for $\tau$ Boo. (b) Single continuum-normalized spectrum of H$\alpha$ line for $\tau$ Boo.}\label{fig1}
\end{figure}

\begin{figure}[h]
 \begin{center}
  \includegraphics[width=68mm]{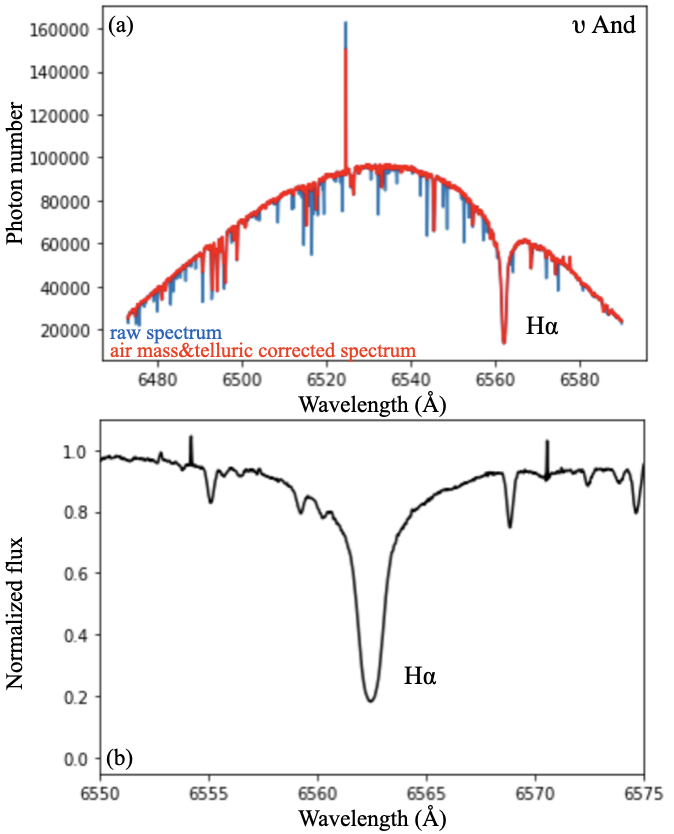}
 \end{center}
 \caption{(a) Single raw spectrum (blue) and telluric line corrected spectrum (red) of H$\alpha$ line for $\upsilon$ And. (b) Single continuum-normalized spectrum of H$\alpha$ line for $\upsilon$ And.}\label{fig2}
\end{figure}

Figure \ref{fig1} (a) shows the single raw spectrum and the telluric line corrected spectrum for $\tau$ Boo obtained after the data reduction. The telluric line corrected spectra throughout the procedure were continuum-normalized by comparing them to a reference spectrum (one single spectrum of May 2, 2019, was chosen) defined as a common flux scale. Figure \ref{fig1} (b) also shows the normalized H$\alpha$ line profile and its core for $\tau$ Boo.
The continuum-normalized spectra were carried out wavelength-shift to be aligned before the analysis. 
Figure \ref{fig2} (a) shows the single raw spectrum and the telluric line corrected spectrum for $\upsilon$ And obtained after the data reduction. The telluric line correction for $\upsilon$ And was carried out in the same manner (one single spectrum of Oct. 19, 2019, was chosen). The normalized H$\alpha$ line profile and its core for $\upsilon$ And are shown in Figure \ref{fig2} (b).

\section{Analyses and results}
\subsection{H$\alpha$ line intensity}
\begin{figure}[t]
 \begin{center}
  \includegraphics[width=80mm]{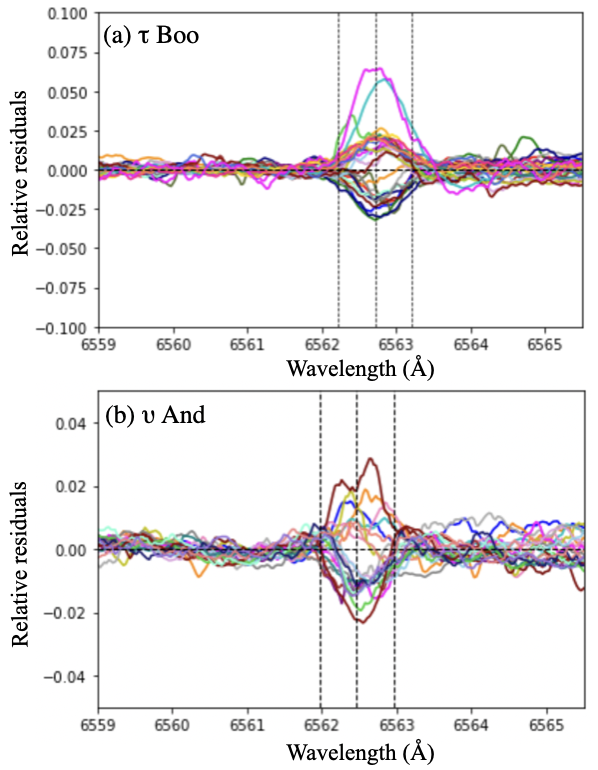}
 \end{center}
 \caption{(a) Residuals relative to the average spectrum in H$\alpha$ lines of $\tau$ Boo (a part of all spectra). (b) Residuals relative to the average spectrum in H$\alpha$ lines of $\upsilon$ And (a part of all spectra). The residuals have been smoothed to avoid cluttering. The dashed lines indicate a range of $\sim$ 1\AA\ around the line core.}\label{fig3}
\end{figure}

To obtain the strength of the variability in the H$\alpha$ line core for each star, we averaged all the continuum-normalized spectra and then calculated the relative residuals for each normalized spectrum with respect to the averaged spectrum.
Figure \ref{fig3} shows the relative residuals of these two stars (a part of all spectra), which have been smoothed to show the variability in the H$\alpha$ line core more clearly.

\begin{figure}
 \begin{center}
  \includegraphics[width=80mm]{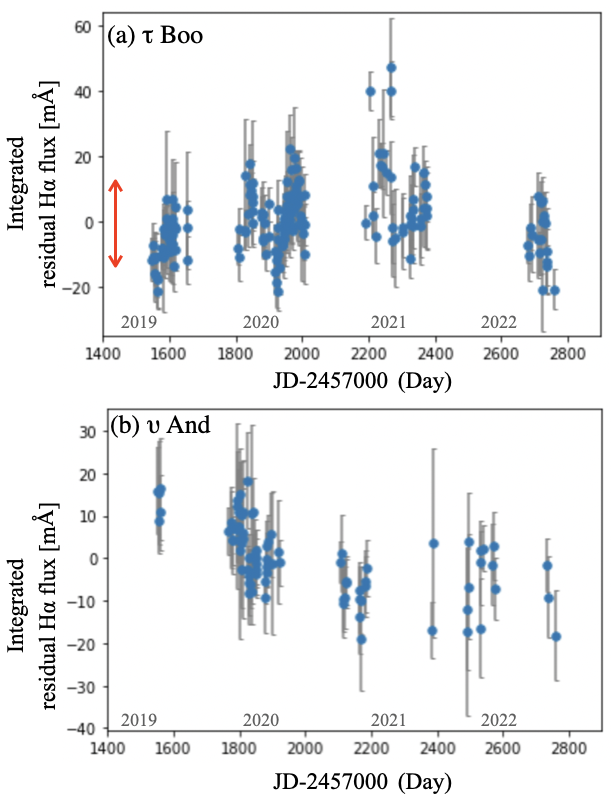}
 \end{center}
 \caption{(a) Integrated residual H$\alpha$ flux of $\tau$ Boo. (b) Integrated residual H$\alpha$ flux of  $\upsilon$ And. The red arrow in the (a) defines the range of the value of \citet{borsa:2015} to compare with our values. The error bars in the integrated residual flux are calculated by the intranight photon noise.}\label{fig4}
\end{figure}

Each star shows a nightly variability in the H$\alpha$ line core throughout the observing seasons. To evaluate the level of the H$\alpha$ variability, we calculated the integrated residual flux. We integrated the relative residuals over a $\sim$ 1\AA\ around the line core. In Figure \ref{fig4}, we plot the integrated residual H$\alpha$ flux [m\AA] along the time series of each star to present the intensity of the H$\alpha$ line core.
The previous H$\alpha$ line observation data for $\tau$ Boo reported by \citet{borsa:2015} would provide great information for comparison of the H$\alpha$ line core intensity, which is obtained from the integral inside a given wavelength range (1 \AA) of the relative residuals. Their data were taken by HARPS-N$/$TNG (\textit{R} $\sim$ 115000, S$/$N $\sim$ 2500-3000), and the integrated residual H$\alpha$ flux lies in a range similar to our values (-10 $\sim$ 10 m\AA). Considering the difference in the signal-to-noise ratio for both instruments and the relatively fewer number of HAPRS-N data, the slightly more extensive values of several HIDES-F data can be explained. Thus, we conclude that the H$\alpha$ flux of HIDES-F observations has a level similar to the previous result and is sufficient to present the line core variability as a stellar magnetic activity tracer. 

\subsection{Long-term variability in the H$\alpha$ line}
 \subsubsection{$\tau$ Boo}
In order to investigate the magnetic activity behavior of $\tau$ Boo, we examined our H$\alpha$ flux time series data. In Figure \ref{fig4} (a), the integrated residual H$\alpha$ flux, especially around 2020, shows a clear periodic variation.

\begin{figure*}[t]
 \begin{center}
  \includegraphics[width=160mm]{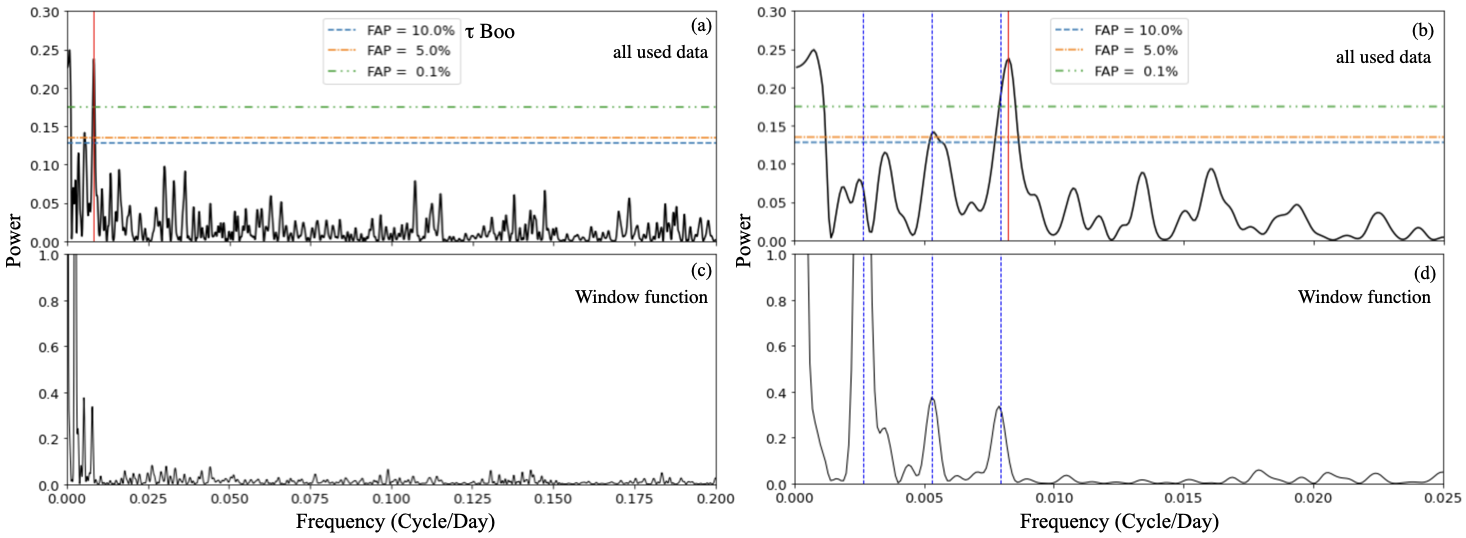}
 \end{center}
 \caption{(a) Periodogram of the integrated residual H$\alpha$ flux time series of $\tau$ Boo. (b) Extended figure of (a). The red line represents the 123 days. (c) Window function of the H$\alpha$ flux time series of $\tau$ Boo. (d) Extended figure of (c). The blue vertical dashed lines represent the aliases of the observation sampling of one year at 1/377 (cycle/day) intervals.}\label{fig5}
\end{figure*}

\begin{figure*}[t]
 \begin{center}
  \includegraphics[clip,height=70mm, width=140mm]{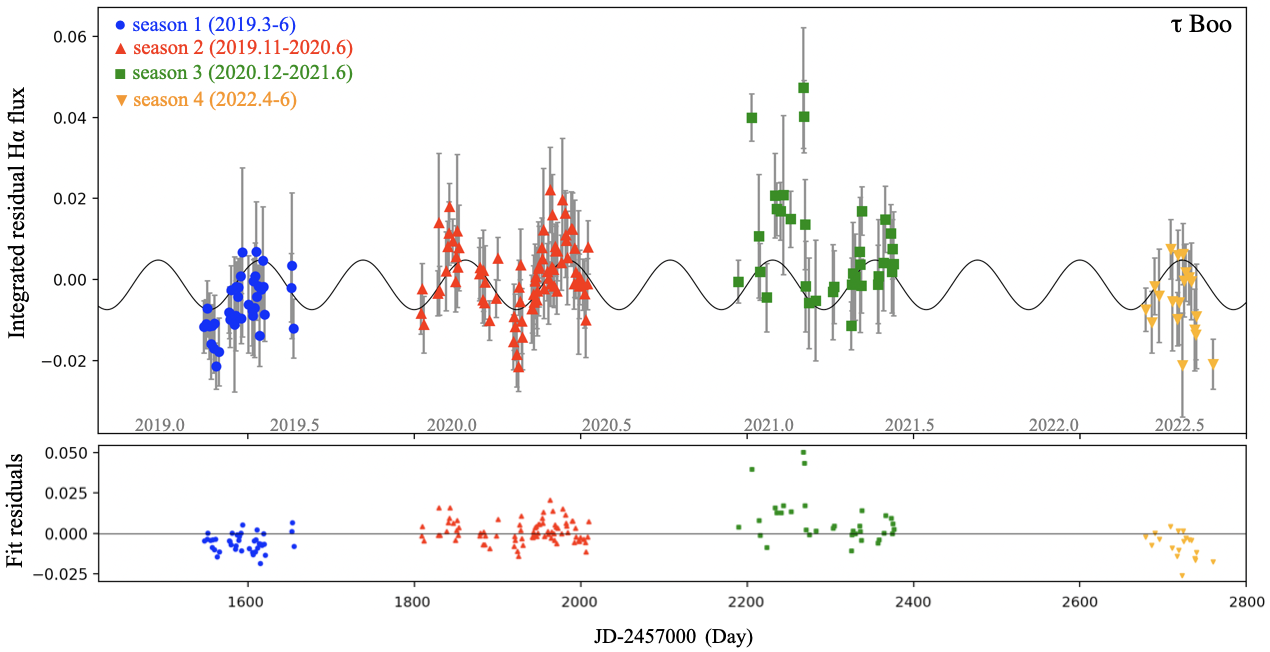}
 \end{center}
 \caption{Integrated residual H$\alpha$ flux time series (upper panel) and the fit residuals (lower panel) of $\tau$ Boo. The solid line depicts the sinusoidal fit with a period of 123 days. The different colors and symbols represent the each observing season during four years (2019-2022). The error bars in the integrated residual flux are calculated by the intranight photon noise.}\label{fig6}
\end{figure*}

To check the period of this distinct variation, we used the generalized Lomb-Scargle (GLS) method by \citet{zechmeister:2009}. The result of the GLS periodogram analysis is shown in Figure \ref{fig5}. A significant peak of the periodogram power was found at 123 days (0.00813 cycle/day) with FAP $<$ 0.1\%. 
We present a sinusoidal curve of the 123 days fitted to the integrated residual H$\alpha$ flux of $\tau$ Boo in Figure \ref{fig6}.
The residual of the integrated residual flux with respect to the fitting sinusoidal curve is also presented (lower panel of Fig. \ref{fig6}), and the fit residuals in 2019 and 2022 show possible increasing or decreasing trends along the time series.

\begin{figure*}[t]
 \begin{center}
  \includegraphics[clip,height=70mm, width=140mm]{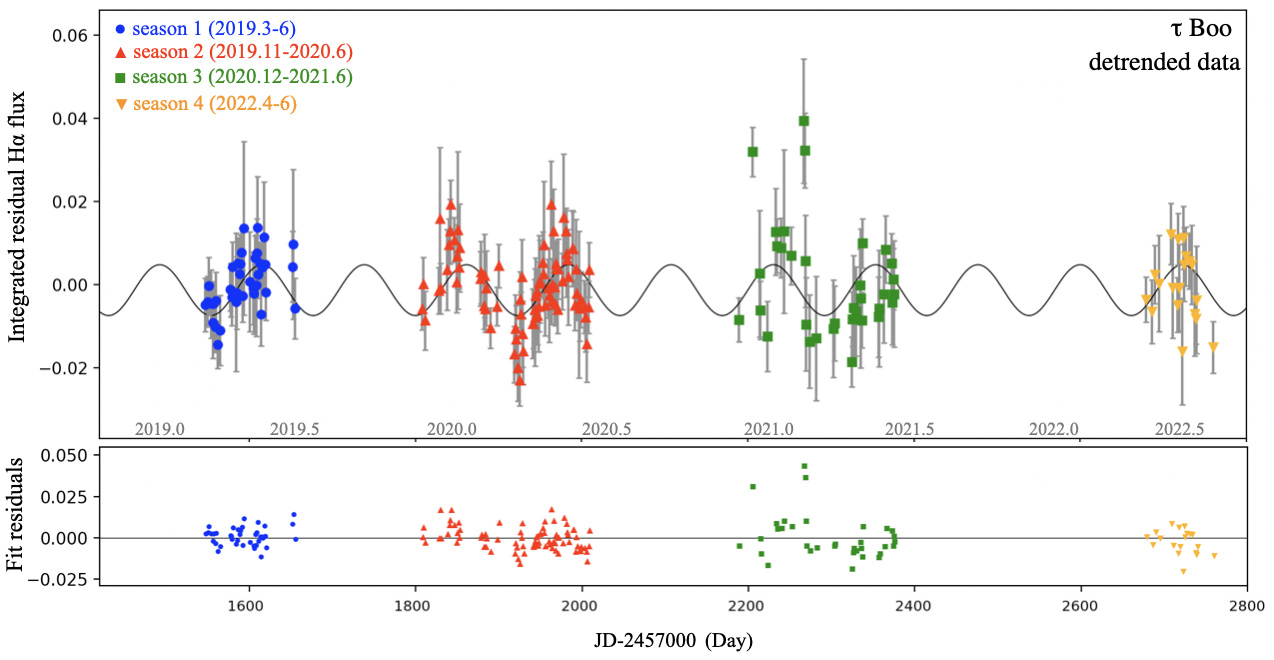}
 \end{center}
 \caption{Detrended data of the integrated residual H$\alpha$ flux time series (upper panel) and the fit residuals (lower panel) of $\tau$ Boo. The solid line depicts the sinusoidal fit with a period of 123 days. The different colors and symbols represent the each observing season during four years (2019-2022). The error bars in the integrated residual flux are calculated by the intranight photon noise.}\label{fig7}
\end{figure*}

\begin{figure*}[t]
 \begin{center}
  \includegraphics[width=160mm]{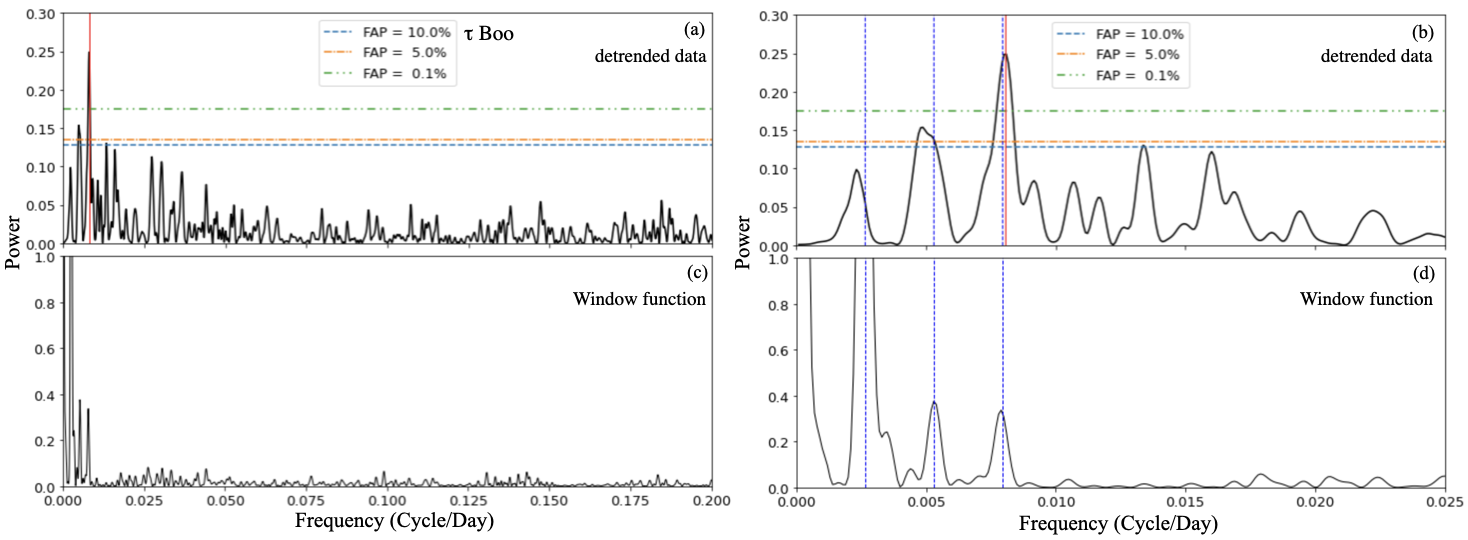}
 \end{center}
 \caption{(a) Periodogram of the detrended H$\alpha$ flux data of $\tau$ Boo. (b) Extended figure of (a). The red line represents the 123-day. (c) Window function of the H$\alpha$ flux time series of $\tau$ Boo. (d) Extended figure of (c). The blue vertical dashed lines represent the aliases of the observation sampling of one year at 1/377 (cycle/day) intervals. }\label{fig8}
\end{figure*}

In the result of the periodogram (Fig. \ref{fig5} (a)\&(b)), a long-term variation ($<$ 0.0008 cycle/day) is also shown with a significant power as well as the 123-day periodicity. 
Its aliases could be represented as an observation sampling of one year with an interval of 1/377 (0.00265) cycle/day according to the window function as shown in Figure 5 (c)\&(d), and one of them is close to the 123-day period. 
However, in the GLS periodogram of the H$\alpha$ flux time series (Fig. 5 (b)), the peak corresponding to the observation sampling, which is located at 377 days (0.00265 cycle/day), is not significant (FAP $>$ 10\%). This is even much smaller than the possible cycle peak at 123 days, although the peak at 377 days should be the highest peak if the observation sampling affects data.
Therefore we suppose that the 123 days is not influenced by the observation sampling of one year.
In order to confirm this, we removed the possible long-term trend ($\approx$ 0.0008 cycle/day) that induces the aliases of an interval of 1/377 cycle/ day (Fig. \ref{fig7}) and calculated the GLS periodogram. As a result, the periodogram of the detrended H$\alpha$ flux clearly shows that the 123 days is still significant with FAP $<$ 0.1\% and the power of the peak is stronger than before the long-term trend was removed (Fig. \ref{fig8}). Thus, we confirmed that the 123-day period is not an alias of a long-term trend but a true signal.

\begin{figure}[t]
 \begin{center}
  \includegraphics[width=100mm]{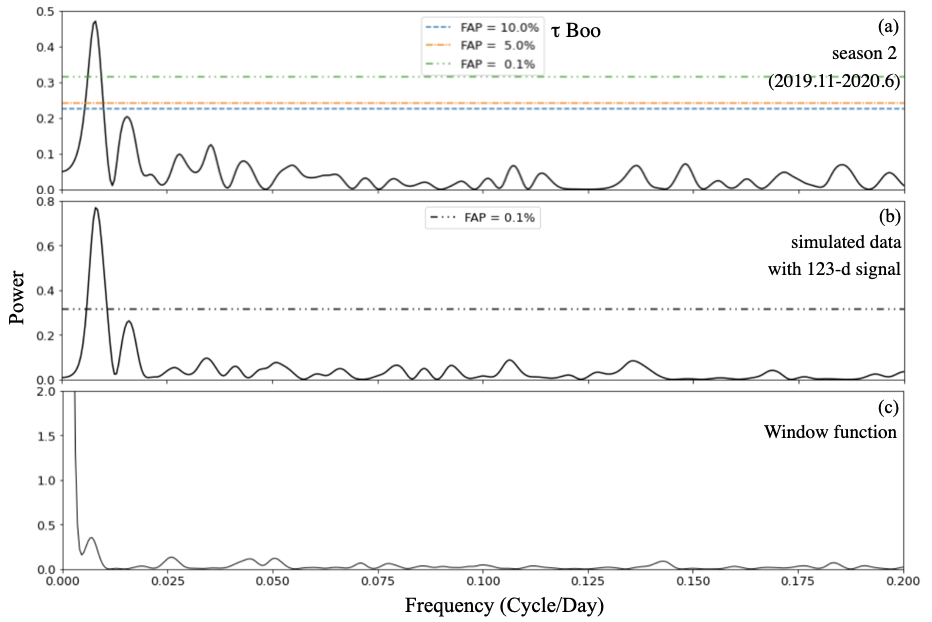}
 \end{center}
 \caption{(a) Periodogram of the integrated residual H$\alpha$ flux time series of the season 2 (2019.11-2020.6) of $\tau$ Boo. (b) Periodogram of the simulated data set with the 123-day period along the integrated residual H$\alpha$ flux time series of the season 2 of $\tau$ Boo. (c) Window function of the H$\alpha$ flux time series of the season 2 of $\tau$ Boo.}\label{fig9}
\end{figure}

We further investigated the 123-day variation focusing on the season 2 (2019.11-2020.6) due to the visible variation with dense time series data within a year. Also, this season does not show a significant long-term trend. We therefore do not need to be concerned about aliases possibly derived from a long-term trend. Figure \ref{fig9} (a) shows the periodogram of the integrated residual H$\alpha$ flux time series in the season 2 of $\tau$ Boo and represents its distinct period at 123 days.

\begin{figure}[t]
 \begin{center}
  \includegraphics[width=80mm]{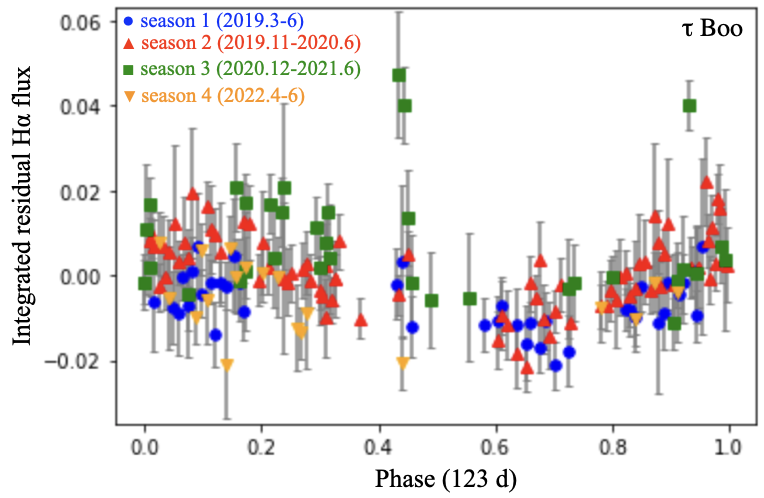}
 \end{center}
 \caption{Phase-folded integrated residual H$\alpha$ flux time series of $\tau$ Boo. The different colors and symbols represent the each observing season during four years (2019-2022). The error bars in the integrated residual flux are calculated by the intranight photon noise.}\label{fig10}
\end{figure}

To further examine the validity of the 123-day periodicity, we adopted so-called a bootstrap randomization method. We generated fake data sets with different periods and amplitudes as a signal of variations along the time series of the season 2. Also, noise data sets were generated by randomly redistributing the fit residuals. We injected the signal of each fake data against the noise, performed the GLS periodogram analysis for each data set, and checked if the 123-day period could appear even when a different period was injected.
As a result, only when we injected a signal with a period of $\sim$ 120 days with an amplitude similar to or greater than observed one, the periodogram showed a peak at $\sim$ 120 days with a FAP well below 0.1\%.
Figure \ref{fig9} (b) shows the periodogram of the simulated data set with the 123-day period, and the window function of the season 2 is also presented (Fig. \ref{fig9} (c)). 
Thus, we conclude that the 123-day variation is independent of a long-term trend and it has been detected as a magnetic activity cycle of $\tau$ Boo in this observation. We show phase-folded H$\alpha$ flux time series of the 123 days in Figure \ref{fig10} as well. 

\begin{figure*}[h]
 \begin{center}
  \includegraphics[width=160mm]{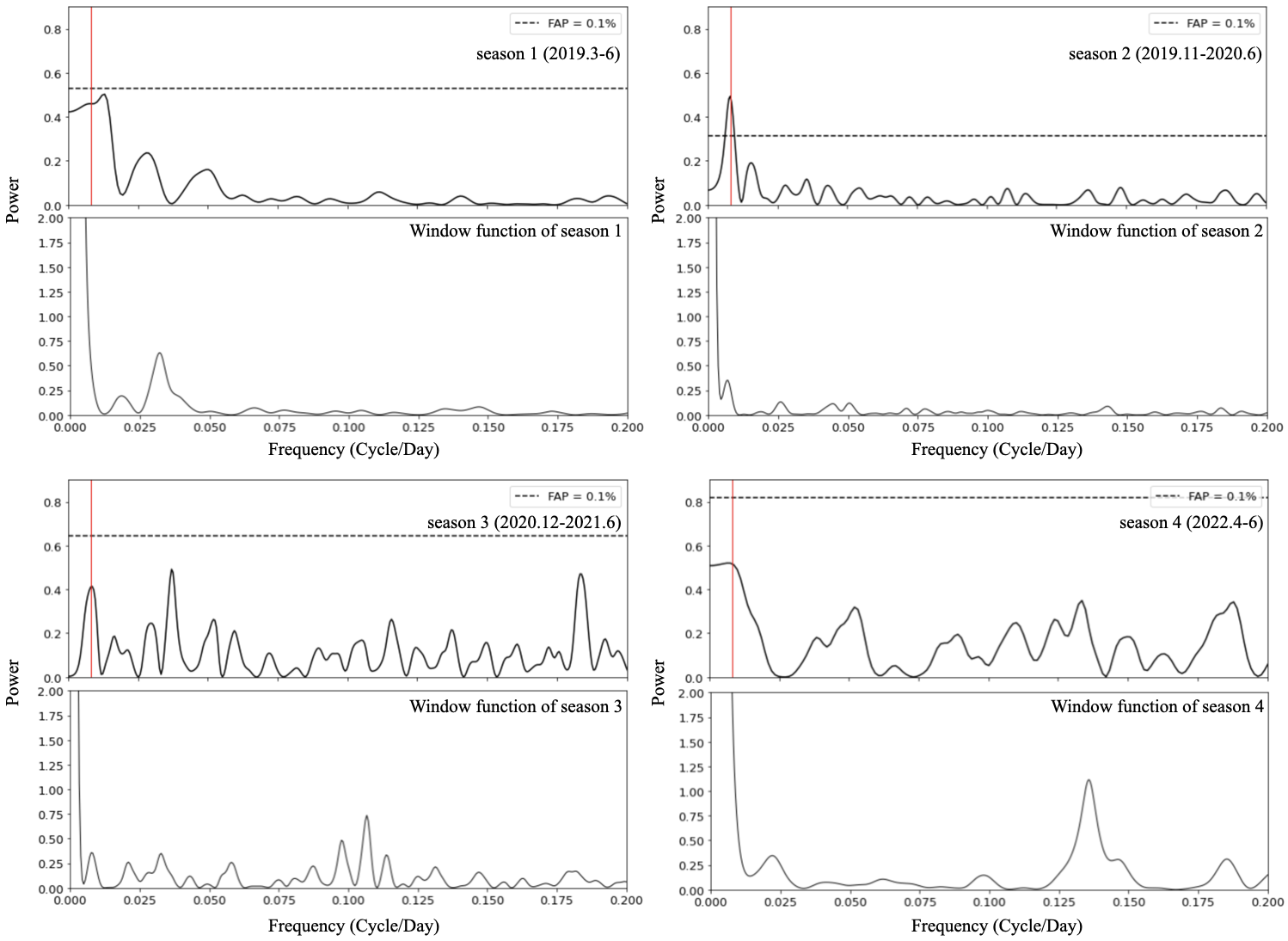}
 \end{center}
 \caption{Periodogram of the seasonal H$\alpha$ flux time series of $\tau$ Boo (upper panel) and window function of the seasonal H$\alpha$ flux time series of $\tau$ Boo (lower panel) for each season. The red lines represent the identified period of 123 days from the result of the entire time series. }\label{fig11}
\end{figure*}

In addition to the result of the seasonal variation in 2020, we analyzed other individual observing seasons separately. The results of the periodogram analysis for each season are shown in Figure \ref{fig11}. 
As described already, the season 2 shows a distinct period of $\approx$ 120 days with a power exceeding much over the FAP=0.1\%. The other three seasons (the season 1, 3, 4) also show a specific period close to $\approx$ 120 days, however, the peaks of the period is presented with a weak power, which could not reach FAP=0.1\%. 
We independently estimated the FAP of the possible peaks by the bootstrap randomization method, in which we randomly redistributed the integrated residual H$\alpha$ flux, calculated the GLS periodogram, obtained the frequency of a period whose power exceeds the highest peak in the periodogram of the H$\alpha$ flux, and repeated the procedure for 10$^{5}$ fake data sets. The frequency thus obtained, that is FAP, is $7 \times {10^{-3}}$ for the season 1, $6 \times {10^{-3}}$ for the season 3, and $2 \times {10^{-2}}$ for the season 4, which are consistent with FAP $>$ 0.1\% estimated by the method in \citet{zechmeister:2009}.
Although the peaks in these seasons are not significant in our data, considering that the periodogram of the entire time series shows the distinct period of $\approx$ 120 days, the existence of $\approx$ 120-day variation in the individual seasons is still implied. The relatively weak periodogram power of the three seasons is possibly due to fewer data, or the temporal absence of the cycle.

The cycle of 123-day in our H$\alpha$ flux data is precisely consistent with the previous 122-day cycle reported by \citet{mittag:2017} using the Ca II H\&K line. This also suggests that the 123-day period represents the magnetic activity cycle on the chromosphere of $\tau$ Boo. 

The time series and the periodogram of the integrated residual H$\alpha$ flux indicate that there might be a more long-term trend beyond approximately 1300 days (0.0008 cycle/day) as well. As the time series of this observation does not cover the possible long-term cycle period, further monitoring to identify a more long-term activity cycle is encouraged.

\subsubsection{$\upsilon$ And}
It has not been discovered that $\upsilon$ And exhibits the magnetic activity cycle on its chromosphere so far. To search a potential activity cycle on $\upsilon$ And, the same analysis as for $\tau$ Boo was carried out. As shown in Figure \ref{fig4} (b), the integrated residual H$\alpha$ flux of $\upsilon$ And is plotted along the time series. 
As we found a visible long-term trend of the integrated residual flux decreasing gradually, we examined whether there is an actual long-term trend in the H$\alpha$ variability with linear or quadratic trends. To address the significance of the trend, we used the F-test in a similar way to that described by \citet{cumming:1999}.
The weighted sum of squares of residuals to the best-fit linear slope $\chi_{N-2}^2$ is compared with the weighted sum of squares about the mean $\chi_{N-1}^2$. If there is no long-term linear trend in the observed H$\alpha$ flux and the residual follow Gaussian distribution, the statistic

\begin{equation}
F = (N-2)\frac{\chi_{N-1}^2-\chi_{N-2}^2}{\chi^{2}_{N-2}}
\end{equation}

\noindent
follows Fisher’s F distribution with 1 and N-2 degrees of freedom which measures how much the fit is improved by introducing the linear trend.
For the quadratic trend, the weighted sum of squares of residuals to the best-fit quadratic curve $\chi_{N-3}^2$ is compared with the weighted sum of squares of residuals to the best-fit linear slope $\chi_{N-2}^2$.

\begin{equation}
F = (N-3)\frac{\chi_{N-2}^2-\chi_{N-3}^2}{\chi^{2}_{N-3}}
\end{equation}

\begin{figure*}[t]
 \begin{center}
  \includegraphics[width=160mm]{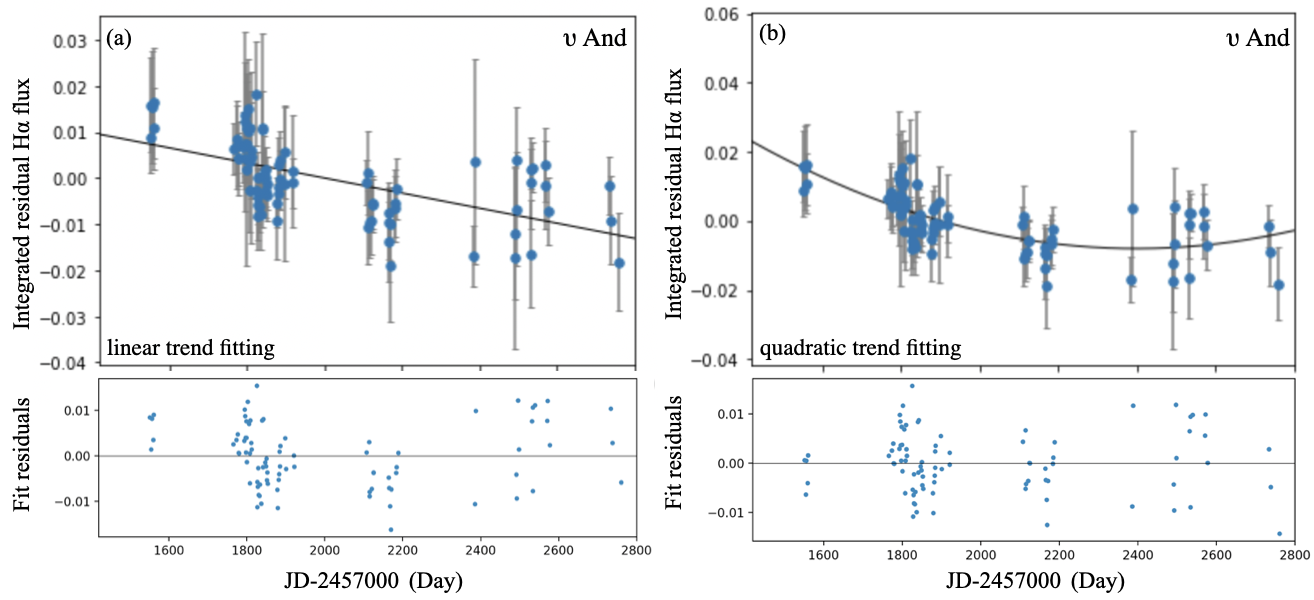}
 \end{center}
 \caption{(a) Integrated residual H$\alpha$ flux time series (upper panel) and the fit residuals (lower panel) with the linear trend for $\upsilon$ And. (b) Integrated residual H$\alpha$ flux time series (upper panel) and the fit residuals (lower panel) with the quadratic trend for $\upsilon$ And. The error bars in the integrated residual flux are calculated by the intranight photon noise.}\label{fig12}
\end{figure*}

\noindent
We also adopted the bootstrap randomization method to estimate FAP that pure noise fluctuations would produce a trend by chance. In this approach, the integrated residual fluxes are randomly redistributed, keeping fixed the observation time. We generated 10$^{5}$ fake data sets in the bootstrap randomization, applied the same analysis to the star, and obtained F-value by the equation (1) or (2) for the linear and quadratic trend case, respectively. The frequency of fake data sets whose F exceeded the observed one, obtained by the equation (1), was adopted as a FAP for the linear trend (\cite{sato:2012}). The FAP for the quadratic trend relative to the linear one indicates the frequency of fake data sets whose F exceeded the observed value obtained by the equation (2).
As a result, both of the FAP were less than 10$^{-5}$, which means that a long-term trend exists in the data and it can be reproduced well by a quadratic trend rather than a linear one (Fig. \ref{fig12}).

\begin{figure*}[t]
 \begin{center}
  \includegraphics[width=160mm]{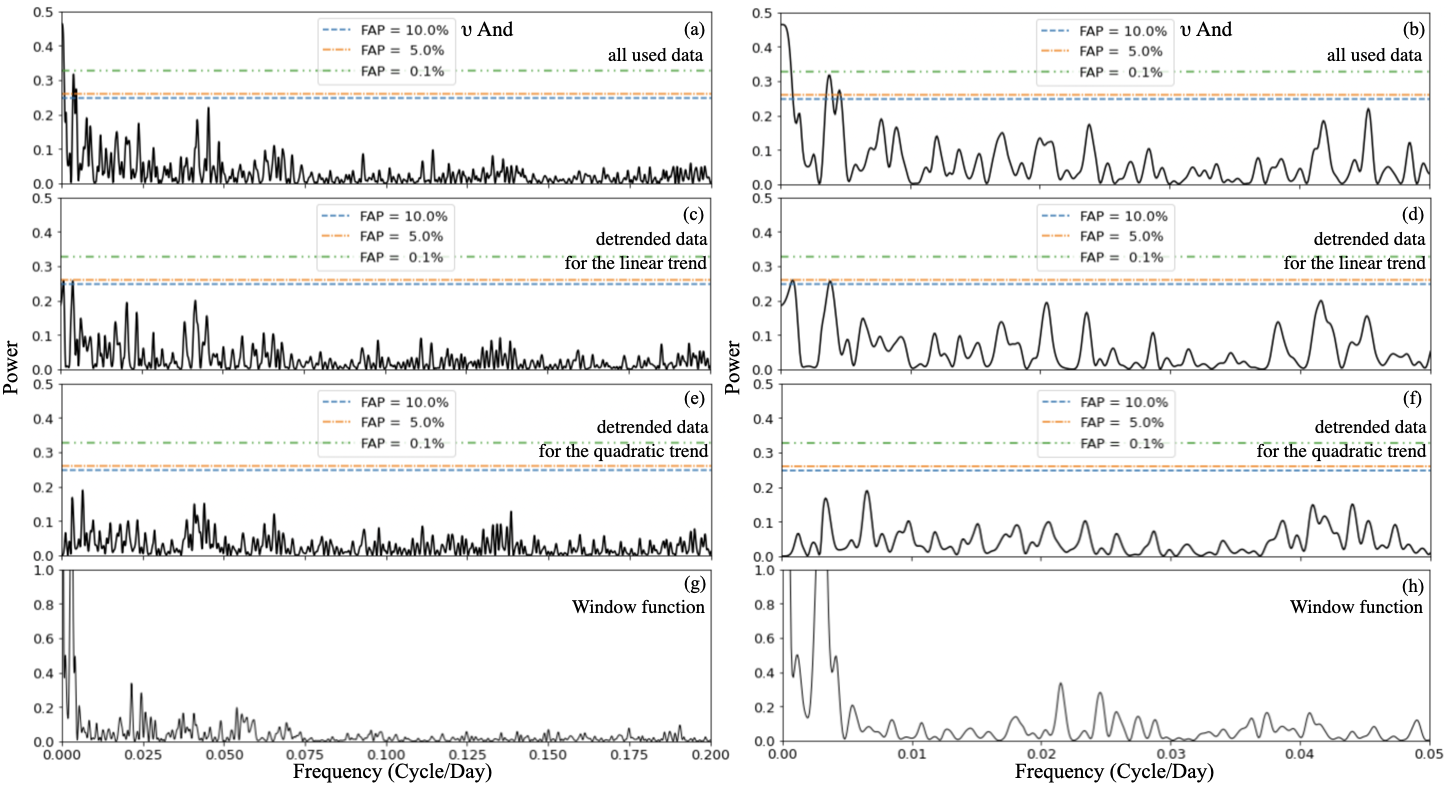}
 \end{center}
 \caption{(a) Periodogram of the integrated residual H$\alpha$ flux time series of $\upsilon$ And. (b) Extended figure of (a). (c) Periodogram of the detrended H$\alpha$ flux data of $\upsilon$ And for the linear trend. (d) Extended figure of (c). (e) Periodogram of the detrended H$\alpha$ flux data of $\upsilon$ And for the quadratic trend. (f) Extended figure of (e). (g) Window function of the H$\alpha$ flux time series of $\upsilon$ And. (h) Extended figure of (g).}\label{fig13}
\end{figure*}

The GLS periodogram of $\upsilon$ And in Figure \ref{fig13} (a)\&(b) shows relatively significant peaks with a power of FAP=0.1\% at 265 days (0.00377 cycle/day) and a more long-term trend. 
To test the independence of the period, we checked changes in the periodogram before and after removing the long-term trend.
For both the linear and the quadratic trend, the peak of the 265-day period is no longer significant after the trend was removed, respectively (Fig. \ref{fig13} (c)\&(d), (e)\&(f)).
Thus, we speculate that the possible 265-day period is influenced by a long-term trend, or there is a potential cycle much longer than the observing time span.

\subsection{Comparison to an inactive star, $\tau$ Cet}
We needed to check that the variations found in these two stars were not affected by instrumental, seasonal nature, or any non-stellar-induced factors.
$\tau$ Cet (HD10700, G8V, V=3.50) is spectrally similar to the Sun and regarded as an inactive star. This star appears so stable that it is expected to show little stellar activity variability. In addition, its activity cycle is reported to be more than 10 years (\cite{choi:2015}). Thus, $\tau$ Cet is an ideal comparison.

\begin{figure}[t]
 \begin{center}
  \includegraphics[width=80mm]{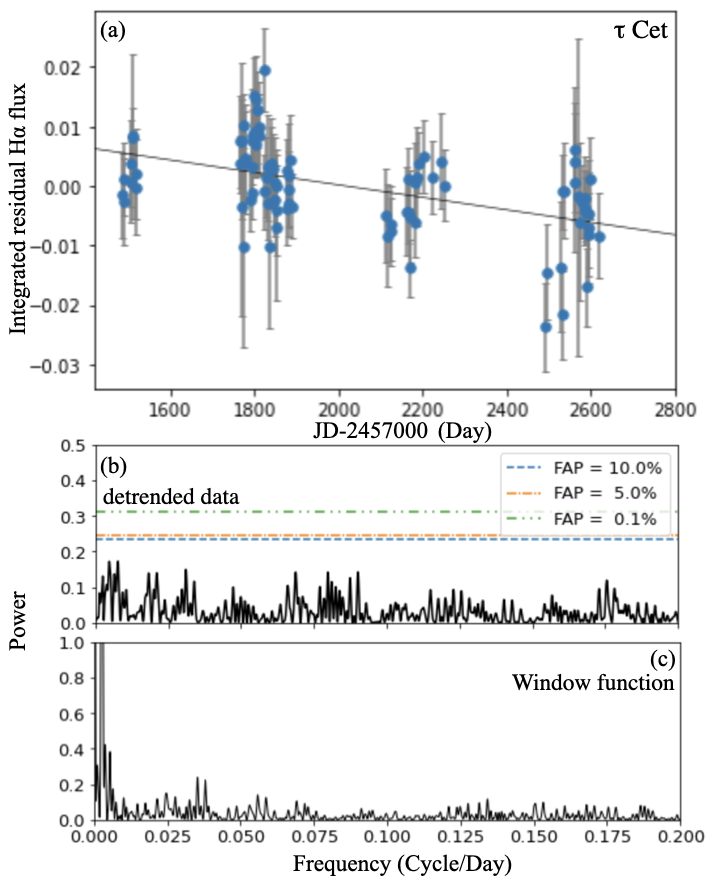}
 \end{center}
 \caption{(a) Integrated residual H$\alpha$ flux time series of $\tau$ Cet with the linear trend. The error bars in the integrated residual flux are calculated by the intranight photon noise. (b) Periodogram of the detrended H$\alpha$ flux data of $\tau$ Cet. (c) Window function of the H$\alpha$ flux time series of $\tau$ Cet.}\label{fig14}
\end{figure}

Figure 14 (a) shows the integrated residual H$\alpha$ flux of $\tau$ Cet observed with HIDES-F in almost the same seasons. To check whether there is a significant trend in the H$\alpha$ variability of $\tau$ Cet, we also performed the linear trend test for this star. As a result, the star showed a linear trend with FAP $<$ 10$^{-5}$. Although the trend might be caused by instrumental, seasonal nature, or stellar intrinsic activity, it is different from that observed in $\upsilon$ And. It suggests the existence of long-term variability in $\upsilon$ And of its own. In order to search for any shorter periodicity, we removed this trend by adjusting the mean integrated residual flux of each season to zero. After detrending, the periodogram of $\tau$ Cet does not show any specific periods such as observed in $\tau$ Boo (Fig. \ref{fig14} (b)). It supports that the observed H$\alpha$ variability of $\tau$ Boo is induced by its intrinsic activity rather than instrumental or seasonal nature.

\subsection{H$\beta$ line variability}
\begin{figure}[]
 \begin{center}
  \includegraphics[width=160mm]{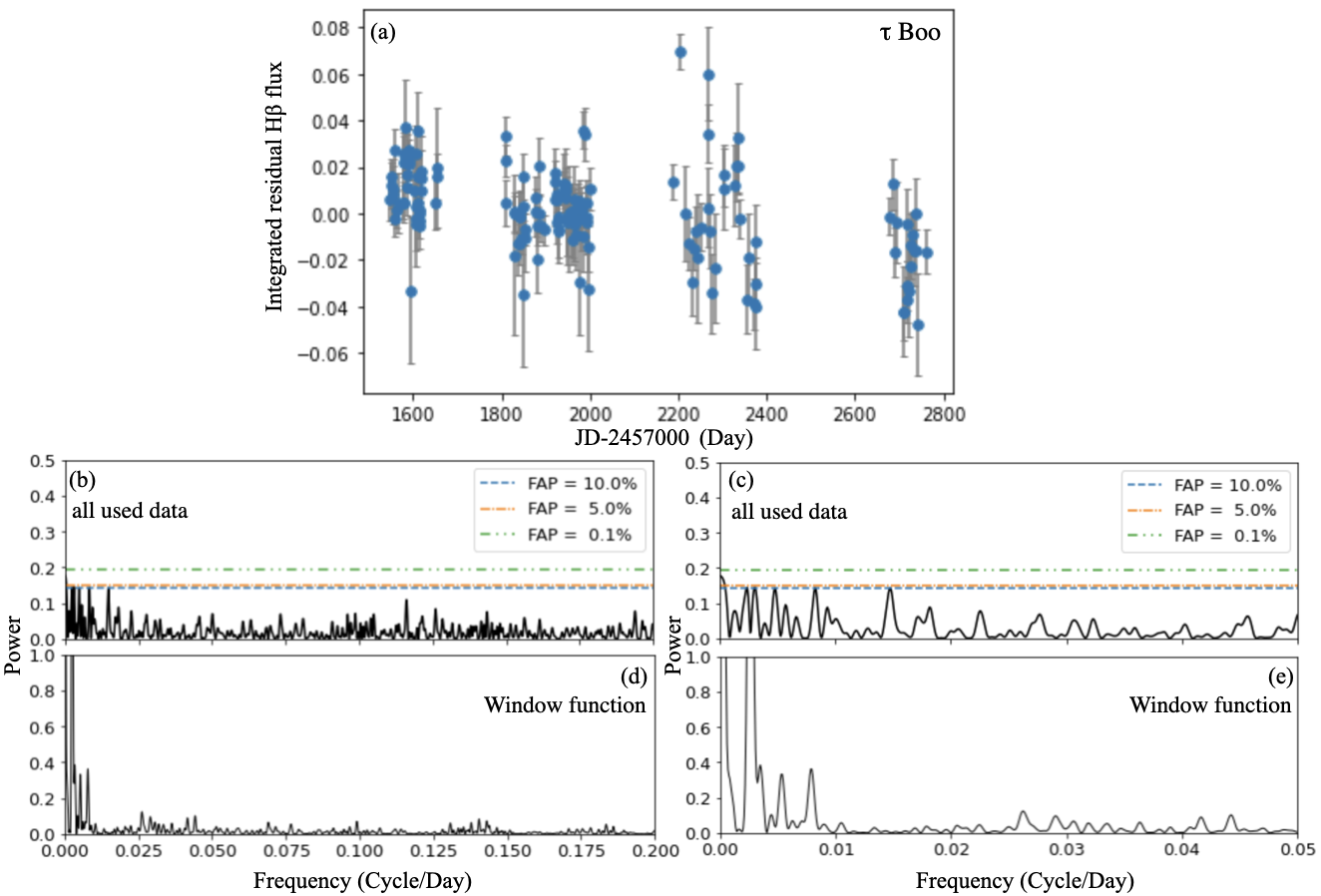}
 \end{center}
 \caption{(a) Integrated residual H$\beta$ flux time series of $\tau$ Boo. The error bars in the integrated residual flux are calculated by the intranight photon noise. (b) Periodogram of the integrated residual H$\beta$ flux time series of $\tau$ Boo. 
 (c) Extended figure of (b).
 (d) Window function of the H$\beta$ flux time series of $\tau$ Boo. (e) Extended figure of (d).}\label{fig15}
\end{figure}

\begin{figure*}[]
 \begin{center}
  \includegraphics[width=160mm]{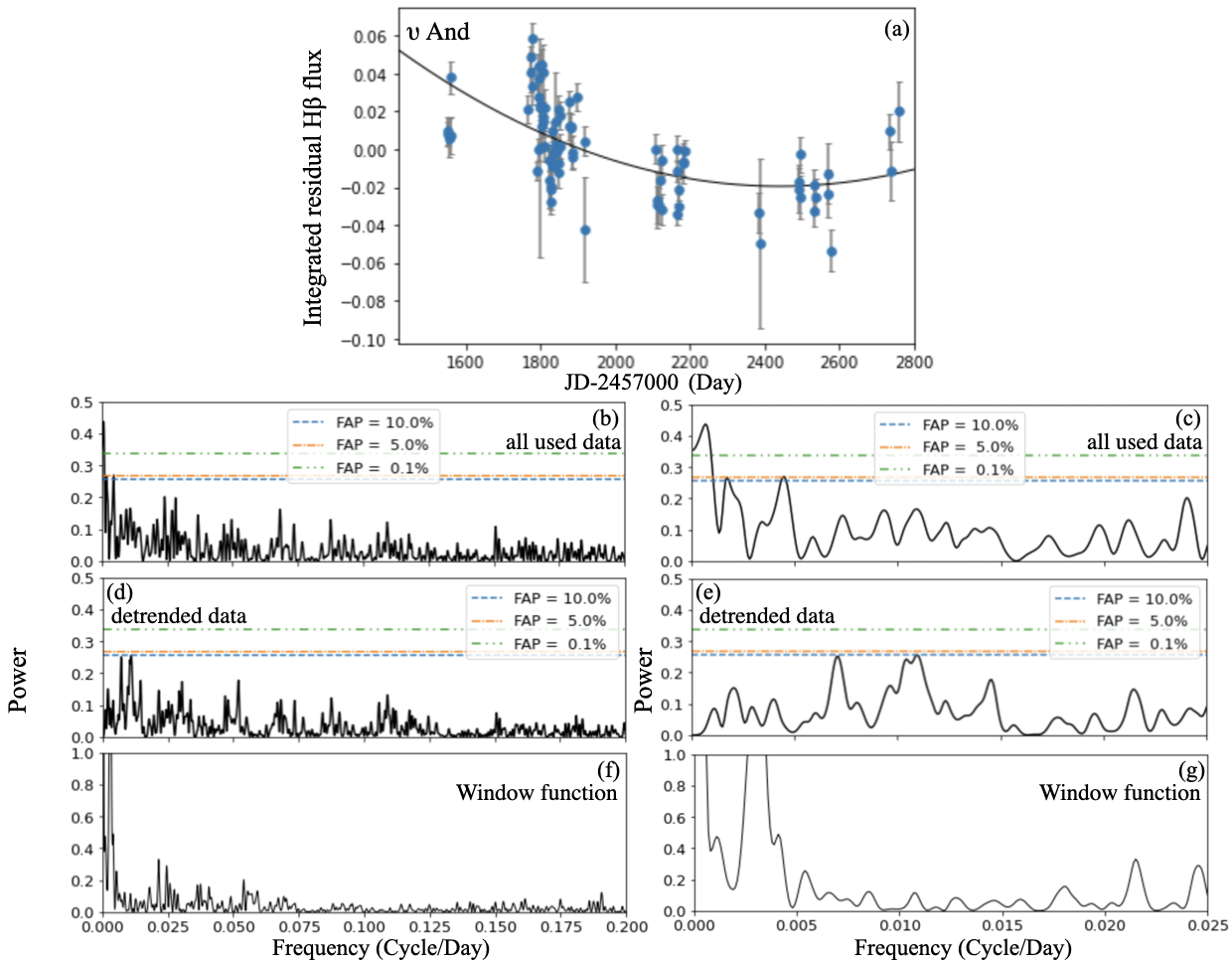}
 \end{center}
 \caption{(a) Integrated residual H$\beta$ flux with the quadratic trend for $\upsilon$ And. The error bars in the integrated residual flux are calculated by the intranight photon noise. (b) Periodogram of the integrated residual H$\beta$ flux time series of $\upsilon$ And. (c) Extended figure of (b).
 (d) Periodogram of the detrended H$\beta$ flux data of $\upsilon$ And. (e) Extended figure of (d).
 (f) Window function of the H$\beta$ flux time series of $\upsilon$ And. (g) Extended figure of (f).}\label{fig16}
\end{figure*}

\begin{figure}[]
 \begin{center}
  \includegraphics[width=80mm]{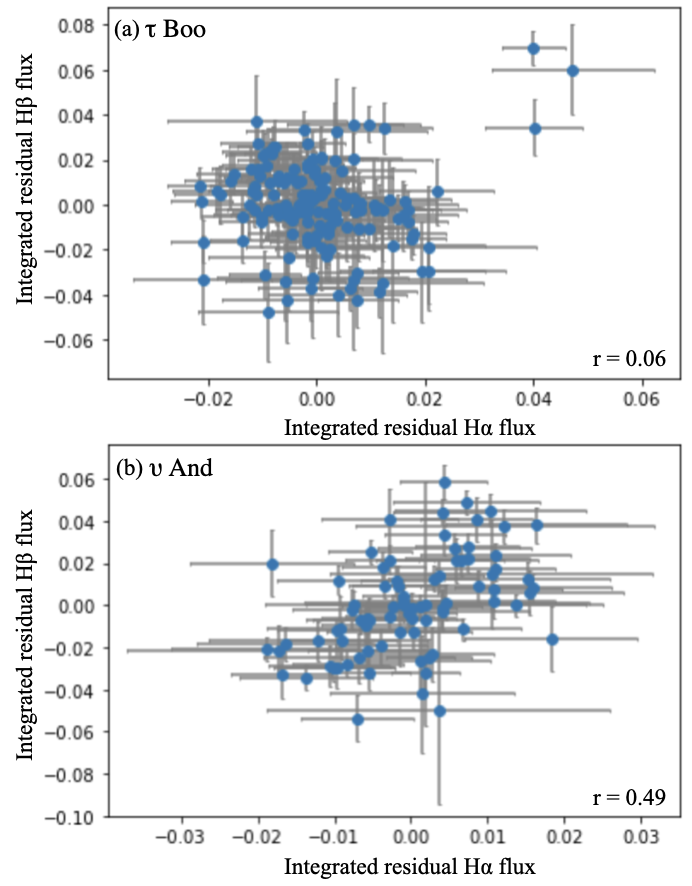}
 \end{center}
 \caption{(a) Correlation between the H$\alpha$ and H$\beta$ flux of $\tau$ Boo. (b) Correlation between the H$\alpha$ and H$\beta$ flux of $\upsilon$ And. The error bars in the integrated residual flux are calculated by the intranight photon noise. The correlation coefficient (r) for each star is described in the figures, respectively.}\label{fig17}
\end{figure}

We also considered using a H$\beta$ line as a part of multi-wavelength observations, to further investigate the activity behaviors of $\tau$ Boo and $\upsilon$ And. The H$\beta$ line core emission (4863 \AA) is observed on the chromosphere so that it is another magnetic activity indicator.
We computed the integrated residual H$\beta$ flux for each star (Fig. \ref{fig15}\&Fig. \ref{fig16}). For $\tau$ Boo, the GLS periodogram analysis showed peaks at around 120 days. However, the power of them is not strong enough to confirm the cycle in the H$\beta$ observations. Moreover, there is no significant correlation between the integrated residual H$\alpha$ and H$\beta$ flux (\textit{r} = 0.06) (Fig. \ref{fig17}). It suggests that the H$\beta$ flux observed this time does not have sufficient strength to show the clear variation of $\tau$ Boo. 

On the other hand, for $\upsilon$ And, the integrated residual H$\beta$ flux shows a quadratic trend along the time series, like its H$\alpha$ result (Fig. \ref{fig16} (a)). 
Furthermore, there is a weak correlation (\textit{r} = 0.49) between these two indices, unlike for $\tau$ Boo (see Fig. \ref{fig17}). In the periodogram, the most significant peak is at $\approx$ 1300 days (0.0008 cycle/day). After the quadratic trend was removed, there was no significant peak in the periodogram (Fig. \ref{fig16} (d)\&(e)).
As a result of combining the periodogram and the correlation, the possibility of a long-term trend in $\upsilon$ And is also inferred from the H$\beta$ observations.

\subsection{Signature of Star-Planet Magnetic Interaction}
\begin{figure}[h]
 \begin{center}
  \includegraphics[width=80mm]{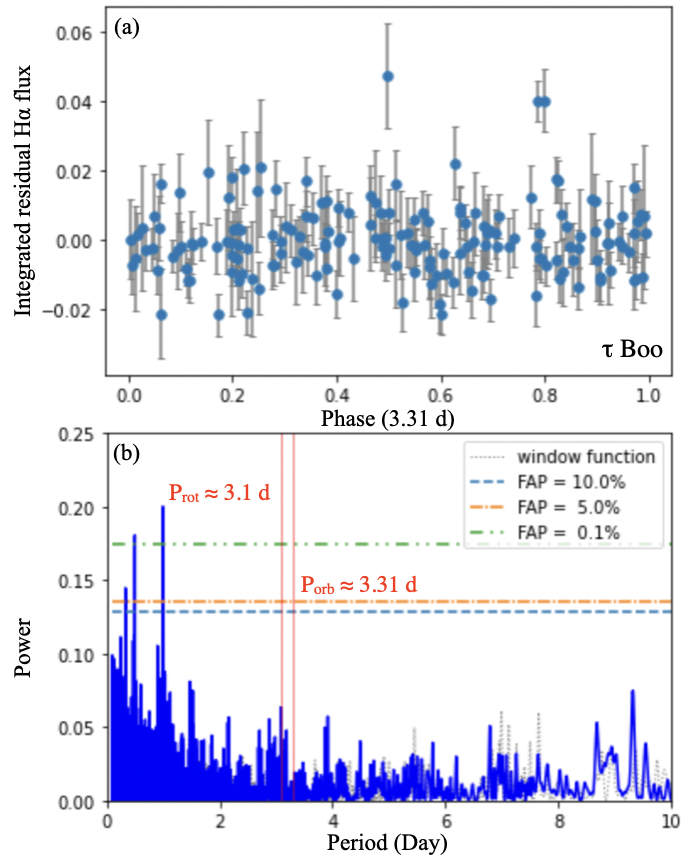}
 \end{center}
 \caption{(a) Phase-folded integrated residual H$\alpha$ flux time series of planetary orbital period for $\tau$ Boo. The error bars in the integrated residual flux are calculated by the intranight photon noise. (b) Periodogram of the integrated residual H$\alpha$ flux time series of $\tau$ Boo. The light-gray dotted line in background shows the window function of the time series. The two red lines show the stellar rotation period and planetary orbital period, respectively.}\label{fig18}
\end{figure}

We monitored the activity variability of $\tau$ Boo and $\upsilon$ And with a focus on hosting a hot-Jupiter and examined any period likely induced by their hot-Jupiter. Given the previous study that has shown the integrated residual flux in the Ca II H\&K line of $\tau$ Boo (\cite{shkolnik:2008}), the H$\alpha$ variability by the SPMI is expected to be below at least 0.2\%. As shown in Figure 6, the H$\alpha$ variability in this observation exceeds 0.2\%. We obtained the H$\alpha$ variability up to 1.4\% as a result of the rms of the amplitude. In this observation data, we could not see a smaller amplitude modulation by the planetary orbital period component when a phase-fold of the orbital period was performed (Fig. \ref{fig18} (a)).  
Also, in the GLS periodogram analysis, a significant peak reaching FAP=0.1\%, which is supposed to be consistent with the planetary orbital period ($\approx$ 3.31 days), has not been detected (Fig. \ref{fig18} (b)). Thus, we suppose that the signature of the SPMI in $\tau$ Boo was not detected in this H$\alpha$ observation. 

\begin{figure}[h]
 \begin{center}
  \includegraphics[width=80mm]{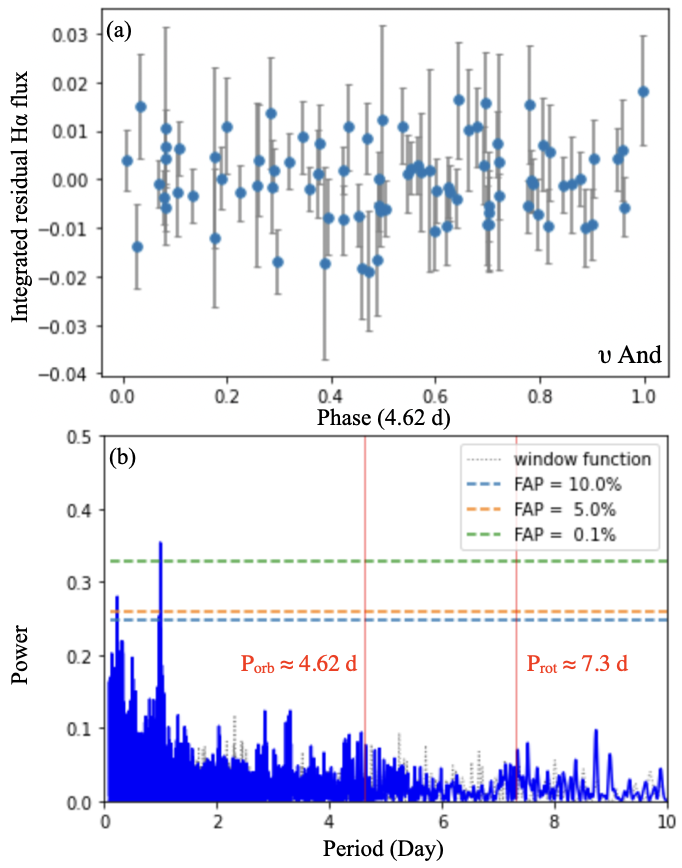}
 \end{center}
 \caption{(a) Phase-folded integrated residual H$\alpha$ flux time series of planetary orbital period for $\upsilon$ And. The error bars in the integrated residual flux are calculated by the intranight photon noise. (b) Periodogram of the integrated residual H$\alpha$ flux time series of $\upsilon$ And. The light-gray dotted line in background shows the window function of the time series. The two red lines show the stellar rotation period and planetary orbital period, respectively.}\label{fig19}
\end{figure}

For $\upsilon$ And, the expected H$\alpha$ variability is at least below 0.5\% given the previous study (\cite{shkolnik:2005}). We obtained the H$\alpha$ variability up to 1.2\% from the rms of the amplitude. Figure \ref{fig19} shows the phase-folded of the orbital period and the periodogram with a range of the orbital period. A significant peak reaching FAP=0.1\%, which is supposed to be consistent with the planetary orbital period ($\approx$ 4.62 days), has not been detected. The non-detection of the SPMI in $\upsilon$ And is being considered in the same way as for $\tau$ Boo.   

Additionally, we removed the tentative long-term trend from the H$\alpha$ variability for both stars, respectively, to investigate any residual variations and their amplitude. However, the amplitude of the residual variations has not changed much, and the periodogram still shows no significant period corresponding to a specific variation induced by the hot-Jupiter for each star.

\section{Discussion}

\subsection{Magnetic activity variability in the H$\alpha$ line}
 \subsubsection{$\tau$ Boo}
Our H$\alpha$ flux time series covers new observing epochs, and its flux is strong enough to trace the activity variability of $\tau$ Boo. We found the existence of a possible $\sim$ 123-day magnetic activity cycle and a more long-term trend in the periodogram (see section 3.2.1). The possible shortest cycle is precisely consistent with the previous Ca II H\&K results. The 122-day cycle, i.e. Four-month cycle, of $\tau$ Boo is expected to have continued for at least the past fifty years through the Mount Wilson project (\cite{baliunas:1997}). Our latest result permits us to further support the long persistence of the four-month cycle in $\tau$ Boo.  

Additionally, our result supports the magnetic activity cycle of 120 days rather than 370 days, among the two values suggested by spectropolarimetry (\cite{fares:2013}). The previous chromospheric results have found a period of four-month consistently. However, as the relationship between the magnetic activity cycle and the large-scale magnetic field cycle in this star might not be similar to that of the Sun, it was still hard to rule out one of the two values (\cite{fares:2013}). Our H$\alpha$ result suggests that 120 days is more likely the magnetic activity cycle by presenting the same chromospheric cycle period with a different indicator than before. Also, our periodogram did not show any significant period at 370 days. The 123-day cycle by dense observations over long-term epochs helps constrain the magnetic field cycle period of $\tau$ Boo.

Also, \citet{mittag:2017} reported a phase-jump event, a shift of phase, in their Ca II H\&K observations during 2013-2016. In the observing season 2015-16, the global fit of 122 days did not provide a perfect fit, and a period of $\approx$ 60 days was found. Also, in that season, an unusual activity behavior was exhibited in which the S-index suddenly dropped from maximum to minimum and increased again, which looks as if the cycle had started a new phase (see Fig. 5 therein). They concluded that this different behavior indicates a shift of phase in the magnetic activity variability of $\tau$ Boo. In addition to the four-month cycle of the star in our H$\alpha$ flux data, we suppose no phase-jump event during the last four years. In our results of the seasonal variation, we did not find unusual activity behaviors along the time series. For the individual periodogram, only the result of the season 2 (2019.11-2020.6) is shown with a strong power. For the other seasons, their periodogram power is not strong as their data numbers might not be sufficient to cover the magnetic activity cycle of the star. 
Nevertheless, it is suggested that a variation close to the four-month cycle is implied in each season (see Fig. \ref{fig11}). More dense time series data of the individual seasons may have exhibited the four-month variation as shown in the entire time series or the season 2.
Given the phase-folded and the global fit of all data, the possibility of any phase-jump before or during this four-year could be low, though we do not rule out a phase-jump event after this time series. Thus, more long time-scale observations covering following time series will be needed.
For the moment, we would like to suggest that the four-month cycle of $\tau$ Boo is possibly persisted without any other phase-jump since the event occurred in 2016.

The Ca II H\&K line and the H$\alpha$ line are formed at the chromosphere, but the depths at which they are formed are slightly different (\cite{vernazza:1981}; \cite{giampapa:1982}). Would these two indicators indeed show this difference through their activity from a perspective of long-term variability? The clear answer and explanation to this question would certainly be a complex of fundamental physics and properties of the atmospheres of the Sun-like stars. We would simply like to emphasize here that a correlation between the two activity indicators is usually accepted (\cite{giampapa:1989}; \cite{walkowicz:2009}; \cite{linsky:2017}). Our H$\alpha$ result showing the same cycle and phase as the latest Ca II H\&K result could defend the correlation in terms of activity behaviors patterns rather than measured activity levels. We therefore interpret the H$\alpha$ line as an efficient tracer as the Ca II H\&K line.

\subsubsection{$\upsilon$ And}
The other F-type star, $\upsilon$ And, has shown the activity variability decreasing gradually along the H$\alpha$ flux time series. We confirmed the existence of its long-term trend with the FAP $<$ 10$^{-5}$ in the linear and the quadratic trend test (see section 3.2.2). Also, the quadratic trend provided a better fit to the time series data of $\upsilon$ And. In the periodogram analysis, the 265 days and a more long-term trend have been found. However, we suggest the possibility that the 265 days is induced by a long-term trend, given that the long-term trend appeared with the most significant power in the periodogram. The 265 days does not appear without this long-term trend.

We could not identify a definite activity cycle in this observation, nevertheless the possibility of a quadratic long-term variation in $\upsilon$ And is suggested. 
Our H$\alpha$ time series in this observation might not be sufficiently long to cover a more long-term variation for this star. Further monitoring to search for a clear activity cycle is encouraged.

\subsection{H$\alpha$ line vs. H$\beta$ line}
In the H$\beta$ observations, we could not find distinct activity cycles for both stars. Although the peak around 123 days is shown in the H$\beta$ result of $\tau$ Boo, we do not classify this result as a detection because it is not sensitive enough to indicate the clear cycle of $\tau$ Boo. In addition, the integrated residual H$\beta$ flux does not seem to be in good agreement with the H$\alpha$ result (see Fig. \ref{fig17}). 

We could see the three data points of both the H$\alpha$ and the H$\beta$ flux increasing rapidly on the same day observation in 2021 (see Fig. \ref{fig6}\&\ref{fig15}). The correlation in Figure \ref{fig17} also shows enhanced fluxes identically in both indicators. This might suggest the possibility of activity enhancements for reasons such as flares. Following-up observations to monitor the enhancements would be interesting.

For $\upsilon$ And, a quadratic long-term trend in the H$\beta$ variability was found. Also, there was no significant peak within the short period range in the periodogram of the detrended H$\beta$ flux data, and it is consistent with the H$\alpha$ result for this star.
Moreover, we found a weak correlation between the H$\alpha$ and the H$\beta$ results (see Fig. \ref{fig17}). The correlation could still support the possibility that there is a long-term trend in $\upsilon$ And.
We have not yet found out why the different behavior appears for these two stars in terms of a correlation between the H$\alpha$ and the H$\beta$ flux. We prefer not to discuss the efficiency of the H$\beta$ line for tracing the magnetic activity at this present. More long-term monitoring in this line for many stars and further research of their atmospheric properties will be needed.

\subsection{Short-term magnetic activity cycle in F-type stars}
\begin{figure}[t]
 \begin{center}
  \includegraphics[width=100mm]{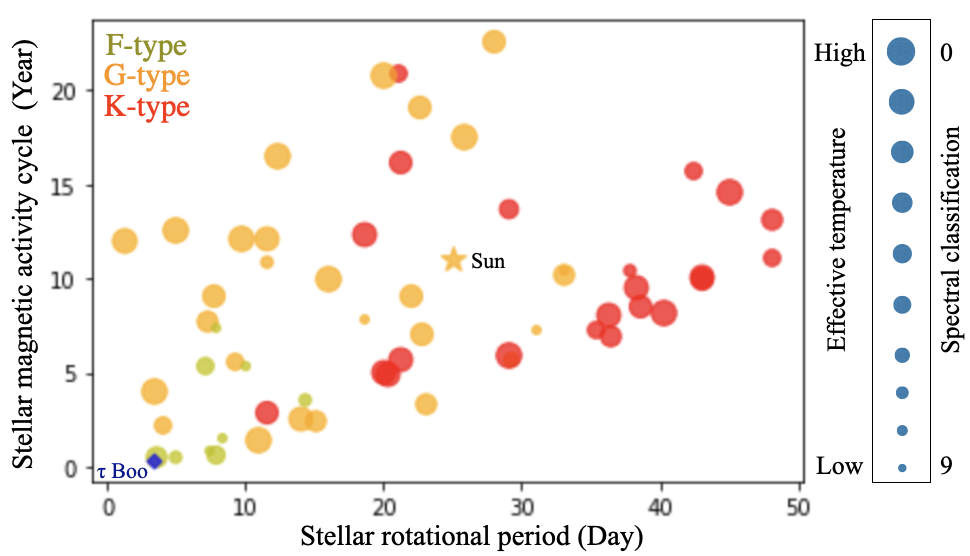}
 \end{center}
 \caption{The magnetic activity cycle period versus the rotational period for F-(light green), G-(yellow) and K-type (red) stars. Points are sized by effective temperature; the bigger point, the earlier type of star. The data are taken from \citet{baliunas:1995}, \citet{saar:1999}, \citet{brandenburg:2017}, \citet{metcalfe:2017} and \citet{mittag:2019}.}\label{fig20}
\end{figure}

The relatively short time-scale of the magnetic activity cycle has been suggested to be a common phenomenon in F-type stars. In addition to the four-month cycle of $\tau$ Boo, a few other F-type stars with a short cycle in the Ca II H\&K observations have been reported (\cite{metcalfe:2010}; \cite{mittag:2019}). 
\citet{metcalfe:2017} suggested that the magnetic activity cycle could be evolved depending on the stellar parameters and showed a relationship between the magnetic activity cycle and the stellar rotational period for the Sun-like stars (see Fig. 3 therein for details).
In this respect, we plot the magnetic activity cycle period versus the rotational period of F-, G- and K-type stars: the data are taken from \citet{baliunas:1995}, \citet{saar:1999}, \citet{brandenburg:2017}, \citet{metcalfe:2017} and \citet{mittag:2019}. Figure \ref{fig20} shows the results for F-, G- and K-type stars, colored light green, yellow and red, respectively. Altogether, it is clear that the faster they rotate, the shorter their magnetic activity cycle is. The magnetic activity cycles of F-type stars appear to be within a range of relatively short-term periods.
Although more cases are needed, it is possible to suggest the short-term magnetic activity cycle trend in F-type stars for the moment.

We observed the persistence of the short-term magnetic activity cycle in the H$\alpha$ variability for $\tau$ Boo (see section 3.2.1), and it is consistent with the previous study (\cite{mittag:2017}).
Our H$\alpha$ results from new observing epochs become an independent confirmation of the short-term magnetic activity cycle of $\tau$ Boo and therefore provide further support for the short-term activity cycle trend in F-type stars. Also, among the F-type stars shown in Figure \ref{fig20}, only $\tau$ Boo has a hot-Jupiter. We found that the possible interaction between $\tau$ Boo and its hot-Jupiter is not strong enough to be detected compared with the stellar magnetic activity cycle of $\tau$ Boo.
Moreover, the short magnetic activity cycle in $\tau$ Boo does not seem to have any relevance to the planetary orbital period (see section 3.5). Thus, in this observation, it is suggested that its hot-Jupiter might not affect the H$\alpha$ variability. We speculate that the short-term magnetic activity cycle of this F-type star is possibly caused by the stellar intrinsic dynamo.

\subsection{Signature of Star-Planet Magnetic Interaction}
Although the SPMI signature has been detected in a few previous Ca II H\&K observations for $\tau$ Boo and $\upsilon$ And (\cite{shkolnik:2005}; \cite{shkolnik:2008}), it still needs a firm conclusion based on different epochs or observational methods. In our H$\alpha$ line monitoring to examine any signal induced by the SPMI, we could not detect a period synchronized with the planetary orbital period. As described in section 3.5, the H$\alpha$ variability expected from the previous research is supposed to be below 0.2\% and 0.5\% for $\tau$ Boo and $\upsilon$ And, respectively. In this observation, we obtained the H$\alpha$ variability of 1.4\% and 1.2\% for each star. Even assuming that smaller amplitude modulations of below 0.2 - 0.5\% occurred, the activity variability is much larger so that it is not easy to detect the possible smaller variability induced by the SPMI. We therefore set the variability of 1.4\% and 1.2\% in the H$\alpha$ line as an upper limit value of non-detection of the SPMI for $\tau$ Boo and $\upsilon$ And, respectively. 

In the latest research about the SPMI of the two stars, \citet{turner:2021} detected radio emission signals that possibly originated from the hot-Jupiter for $\tau$ Boo. They assumed this could be a detection of the SPMI at the time. However, we regard the results of our H$\alpha$ line as the latest non-detection of the SPMI signature. It is speculated that the activity variability is much stronger than the variability possibly induced by the SPMI. Thus, it is not easy to distinguish the possible SPMI signature in this observation. There could also be another reason, such as the precision of the H$\alpha$ observations being insufficient.

To confirm the origin and nature of the activity variability clearly, searching for periodicity and explaining how it relates to possible sources (e.g. stellar activity variability, planetary motion) through various observations in different epochs are essential. In this context, we find the significance of our results as the latest variability related to the stellar activity variability of $\tau$ Boo and $\upsilon$ And. Future investigations of these two stars from various aspects, including the possibility of relevance between the radio emissions signal and the activity variability, will be required to clarify the existence of the SPMI.

\section{Summary}
We investigated the magnetic activity variability for two F-type stars having a hot-Jupiter, $\tau$ Boo and  $\upsilon$ And, by monitoring HIDES-F H$\alpha$ line during the last four years.
As a result, these two stars have shown their intrinsic magnetic activity variability, respectively.

We suggest $\sim$ 123-day magnetic activity cycle of $\tau$ Boo in our H$\alpha$ flux data by the periodogram analysis of the simulated data indicating that the 123-day period represents a real variation. As a part of multi-wavelength observations, we found an identical pattern of activity behaviors with the previous Ca II H\&K results. We suggest that the H$\alpha$ line could be effectively used to perform a survey for the stellar magnetic activity and its cycle. Also, our result allows us to constrain the magnetic field cycle period of $\tau$ Boo.

In addition, we found a long-term trend in $\upsilon$ And, although a clear activity cycle period was not detected in this observation. The quadratic long-term trend in the H$\alpha$ variability might imply one of magnetic activity cycles undiscovered yet in this star.

In the monitoring focusing on a hot-Jupiter, any periods induced by their hot-Jupiter have not been detected for both stars. We set the H$\alpha$ variability of 1.4\% and 1.2\% as an upper limit value of the SPMI non-detection for $\tau$ Boo and $\upsilon$ And, respectively. Further observations to clarify the SPMI of these two stars are still required.
This result could also support that the short-term magnetic activity cycle detected in $\tau$ Boo is not likely related to its hot-Jupiter, but rather to the stellar intrinsic dynamo in the thin outer convective zone. The short time-scale cycle in F-type stars would be an attractive constraint condition for studying the dynamo of Sun-like stars.

 \begin{ack}
This research is based on data collected with 1.88-m telescope at Okayama Branch Office, Subaru Telescope. The Okayama 1.88-m telescope is operated by a consortium led by Exoplanet Observation Research Center, Tokyo Institute of Technology (Tokyo Tech), under the framework of tripartite cooperation among Asakuchi-city, NAOJ, and Tokyo Tech from 2018. B.S. was partly supported by Grant-in-Aid for Scientific Research on Innovative Areas 18H05442 from the Japan Society for the Promotion of Science (JSPS), and by Satellite Research in 2017-2020 from Astrobiology Center, NINS. Y.N. is supported by Grant-in-Aid for KAKENHI Grant Number 21J00106 from JSPS.

 \end{ack}

\end{document}